\documentclass[reprint,prl,twocolumn,superscriptaddress,amsmath,amssymb]{revtex4-2}
\UseRawInputEncoding
\usepackage{graphicx}
\usepackage{multirow}
\usepackage{CJK}

\usepackage[colorlinks,linkcolor=blue,anchorcolor=blue,citecolor=blue]{hyperref}
\usepackage{epstopdf}

\usepackage{color}

\begin{document}
%\begin{CJK*}{GBK}{kai}

\title{New $\alpha$-Emitting Isotope $^{214}$U and Abnormal Enhancement of $\alpha$-Particle Clustering in Lightest Uranium Isotopes}

\author{Z.~Y.~Zhang}
\affiliation{CAS Key Laboratory of High Precision Nuclear Spectroscopy, Institute of Modern Physics, Chinese Academy of Sciences, Lanzhou 730000, China}
\affiliation{School of Nuclear Science and Technology£¬University of Chinese Academy of Sciences, Beijing 100049, China}
\author{H.~B.~Yang}
\affiliation{CAS Key Laboratory of High Precision Nuclear Spectroscopy, Institute of Modern Physics, Chinese Academy of Sciences, Lanzhou 730000, China}
\author{M.~H.~Huang}
\author{Z.~G.~Gan} \email[Corresponding author: ]{zggan@impcas.ac.cn}
\affiliation{CAS Key Laboratory of High Precision Nuclear Spectroscopy, Institute of Modern Physics, Chinese Academy of Sciences, Lanzhou 730000, China}
\affiliation{School of Nuclear Science and Technology£¬University of Chinese Academy of Sciences, Beijing 100049, China}
\author{C.~X.~Yuan}
\affiliation{Sino-French Institute of Nuclear Engineering and Technology, Sun Yat-Sen University, Zhuhai 519082, China}
\author{C.~Qi}
\affiliation{Department of Physics, Royal Institute of Technology (KTH), Stockholm SE-10691, Sweden}
\author{A.~N.~Andreyev}
\affiliation{Department of Physics, University of York, York, YO10 5DD, United Kingdom}
\affiliation{Advanced Science Research Center, Japan Atomic Energy Agency, Tokai, Ibaraki 319-1195, Japan}
\author{M.~L.~Liu}
\affiliation{CAS Key Laboratory of High Precision Nuclear Spectroscopy, Institute of Modern Physics, Chinese Academy of Sciences, Lanzhou 730000, China}
\affiliation{School of Nuclear Science and Technology£¬University of Chinese Academy of Sciences, Beijing 100049, China}

\author{L.~Ma}
\author{M.~M.~Zhang}
\author{Y.~L.~Tian}
\affiliation{CAS Key Laboratory of High Precision Nuclear Spectroscopy, Institute of Modern Physics, Chinese Academy of Sciences, Lanzhou 730000, China}
\author{Y.~S.~Wang}
\affiliation{CAS Key Laboratory of High Precision Nuclear Spectroscopy, Institute of Modern Physics, Chinese Academy of Sciences, Lanzhou 730000, China}
\affiliation{School of Nuclear Science and Technology£¬University of Chinese Academy of Sciences, Beijing 100049, China}
\affiliation{School of Nuclear Science and Technology£¬Lanzhou University, Lanzhou 730000, China}

\author{J.~G.~Wang}
\author{C.~L.~Yang}
\author{G.~S.~Li}
\author{Y.~H.~Qiang}
\author{W.~Q.~Yang}
\author{R.~F.~Chen}
\author{H.~B.~Zhang}
\author{Z.~W.~Lu}
\affiliation{CAS Key Laboratory of High Precision Nuclear Spectroscopy, Institute of Modern Physics, Chinese Academy of Sciences, Lanzhou 730000, China}

\author{X.~X.~Xu}
\author{L.~M.~Duan}
\author{H.~R.~Yang}
\author{W.~X.~Huang}
\author{Z.~Liu}
\author{X.~H.~Zhou}
\author{Y.~H.~Zhang}
\author{H.~S.~Xu}
\affiliation{CAS Key Laboratory of High Precision Nuclear Spectroscopy, Institute of Modern Physics, Chinese Academy of Sciences, Lanzhou 730000, China}
\affiliation{School of Nuclear Science and Technology£¬University of Chinese Academy of Sciences, Beijing 100049, China}

\author{N.~Wang}
\author{H.~B.~Zhou}
\author{X.~J.~Wen}
\author{S.~Huang}
\affiliation{Guangxi Key Laboratory of Nuclear Physics and Technology, Guangxi Normal University, Guilin 541004, China}

\author{W.~Hua}
\author{L.~Zhu}
\affiliation{Sino-French Institute of Nuclear Engineering and Technology, Sun Yat-Sen University, Zhuhai 519082, China}

\author{X.~Wang}
\affiliation{State Key Laboratory of Nuclear Physics and Technology, School of Physics, Peking University, Beijing 100871, China}

\author{Y.~C.~Mao}
\affiliation{Department of Physics, Liaoning Normal University, Dalian 116029, China}

\author{X.~T.~He}
\affiliation{College of Material Science and Technology, Nanjing University of Aeronautics and Astronautics, Nanjing 210016, China}

\author{S.~Y.~Wang}
\author{W.~Z.~Xu}
\author{H.~W.~Li}
\affiliation{Shandong Provincial Key Laboratory of Optical Astronomy and Solar-Terrestrial Environment, School of Space Science and Physics, Institute of Space Sciences, Shandong University, Weihai 264209, China}

\author{Z.~Z.~Ren}
\affiliation{School of Physics Science and Engineering, Tongji University, Shanghai 200092, China}

\author{S.~G.~Zhou}
\affiliation{CAS Key Laboratory of Theoretical Physics, Institute of Theoretical Physics, Chinese Academy of Sciences, Beijing 100190, China}
\affiliation{Center of Theoretical Nuclear Physics, National Laboratory of Heavy-Ion Accelerator, Lanzhou 730000, China}

\begin{abstract}
  A new $\alpha$-emitting isotope $^{214}$U, produced by fusion-evaporation reaction $^{182}$W($^{36}$Ar, 4n)$^{214}$U, was identified by employing the gas-filled recoil separator SHANS and recoil-$\alpha$ correlation technique. More precise $\alpha$-decay properties of even-even nuclei $^{216,218}$U were also measured in reactions of $^{40}$Ar, $^{40}$Ca with $^{180, 182, 184}$W targets. By combining the experimental data, improved $\alpha$-decay reduced widths $\delta^2$ for the even-even Po--Pu nuclei in the vicinity of magic neutron number $N=126$ were deduced. Their systematic trends are discussed in terms of $N_{p}N_{n}$ scheme in order to study the influence of proton-neutron interaction on $\alpha$ decay in this region of nuclei. It is strikingly found that the reduced widths of $^{214,216}$U are significantly enhanced by a factor of two as compared with the $N_{p}N_{n}$ systematics for the $84 \leq Z \leq 90$ and $N<126$ even-even nuclei. The abnormal enhancement is interpreted by the strong monopole interaction between the valence protons and neutrons occupying the $\pi 1f_{7/2}$ and $\nu 1f_{5/2}$ spin-orbit partner orbits, which is supported by a large-scale shell model calculation.
\end{abstract}

\maketitle
%\end{CJK*}

%%%%%%%%%%%%%%%%%%%%%%%%%%%%%%%%%%%%%%%%%%%%%%%%%%%%%%%%%%%%%%%%%%%%%%%%%%%%%%%%%%%%%%%%%%%%%%%%%%%%%%%%%%%%%%%%%%%%%%%%%%%
%\section{Introduction}
%%%%%%%%%%%%%%%%%%%%%%%%%%%%%%%%%%%%%%%%%%%%%%%%%%%%%%%%%%%%%%%%%%%%%%%%%%%%%%%%%%%%%%%%%%%%%%%%%%%%%%%%%%%%%%%%%%%%%%%%%%%

Nucleon-nucleon interaction, which governs the existence of nuclear system, plays a fundamental role in understanding of the properties of exotic nuclei far from stability. Although the proton-proton ($p$-$p$) and neutron-neutron ($n$-$n$) correlations are well-known to be crucial for explaining a wealth of experimental data, the proton-neutron ($p$-$n$) interaction has long been recognized as one of the essential driving forces for the shell structure evolution, the development of collectivity and the onset of deformation in atomic nuclei \cite{Casten2001, Casten2016, Sorlin2008, Otsuka2020, Faestermann2013, Frauendorf2014, Qi2016, Cederwall2011, Chen2009, Neidherr2009}. In the last decades, thanks to the development of radioactive beam facilities worldwide, enormous progress in the physics of change of nuclear shell structure as a function of proton and/or neutron numbers has been achieved in light nuclei. However, the experimental knowledge for the structure evolution in heavy nuclei around and below the neutron closed shell at $N=126$ remains scarce at present \cite{Sorlin2008, Andreyev2013, Khuyagbaatar2015, Hauschild2001, Zhang2019, Ma2020}.

It is well known that, because of large overlap of the radial wave functions, the attractive and short-range interaction between valence protons and neutrons occupying orbits with the same number of nodes and orbital angular momenta (i.e., $\Delta n = \Delta l=0$) becomes stronger than those in other categories, and eventually triggers the remarkable changes of closed shells (see \cite{Sorlin2008, Otsuka2020}, and references therein). For instance, the $\pi 0f_{7/2}$--$\nu 0f_{5/2}$ interaction was shown to play an important role for structure evolution of $N=34$ isotones of Ca, Ti, Cr, and Fe, culminating in the creation of new magic numbers at $N=32$, 34 in $^{52,54}$Ca (see Fig.~1 from Ref.~\cite{Steppenbeck2013} and Ref.~\cite{Wienholtz2013}). In the trans-lead nuclear region with $Z>82$ and $N \leq 126$, the valence protons fill the $0h_{9/2}$, $1f_{7/2}$, and $0i_{13/2}$ orbits, while the neutrons mainly occupy the $2p_{1/2}$, $1f_{5/2}$, and $2p_{3/2}$ orbits \cite{Caurier2003, Teruya2016}. Therefore, one can expect the monopole $p$-$n$ interaction between the $\pi 1f_{7/2}$ and $\nu 1f_{5/2}$ spin-orbit partner orbits to have a significant impact on nuclear structure evolution in that region.

$\alpha$-decay spectroscopy has been proven to be a powerful tool to probe the nuclear structure in heavy nuclei \cite{Duppen2018, Qi2019, Delion2018, Ni2010}. There are analytical formulae to calculate the $\alpha$-decay half-lives such as the new Geiger-Nuttall law \cite{Ni2008, Ren2012}. Typically, the $\alpha$-decay process is described by the two-step mechanism, involving the preformation of $\alpha$ particle followed by its penetration through Coulomb and centrifugal barriers. The $\alpha$-particle preformation probability involves all the nuclear structure information, and can be weighed experimentally by $\alpha$-decay reduced width $\delta^2$ \cite{Rasmussen1959} or the model-independent formation probability $|R\mathcal{F}_{\alpha}(R)|^2$ \cite{Qi2009a, Qi2009b}. It is interesting to note that the $\alpha$-decay reduced widths of several $Z \sim N$ nuclei around $^{100}$Sn ($Z=N=50$) are enhanced by at least a factor of two relative to the benchmark nucleus $^{212}_{~84}$Po$_{128}$ and its neighbouring Po isotopes \cite{Liddick2006, Darby2010, Seweryniak2006, Auranen2018}. This enhancement was explained by the so-called ``superallowed $\alpha$ decay'' \cite{Duppen2018, Macfarlane1965} in relation to the fact that the valence protons and neutrons are in the same single-particle levels, giving rise to a strong $p$-$n$ interaction. In fact, the influence of $p$-$n$ interaction on the absolute $\alpha$-decay widths in $^{212}$Po and nearby nuclei was usually neglected in microscopic calculation, since the low-lying proton and neutron single-particle states are very different from each other in these cases \cite{Tonozuka1979, Qi2010, Qi2019}. However, several theoretical treatments \cite{DodigCrnkovic1985, Varga1992, Lovas1998} pointed to the particular significance of $p$-$n$ interaction in $\alpha$ decay for these nuclei.

In this Letter, we report on the observation of a new isotope $^{214}_{~92}$U$_{122}$ and on more precise measurements for the $\alpha$-decay properties of $^{216,218}$U ($N=124,126$). In this region of nuclei, the protons and neutrons can occupy the $\pi 1f_{7/2}$ and $\nu 1f_{5/2}$ spin-orbit partner orbits to a large extent. Thus, such nuclei can provide a unique opportunity to test the influence of $p$-$n$ interaction on $\alpha$-particle clustering in heavy nuclear region.

%%%%%%%%%%%%%%%%%%%%%%%%%%%%%%%%%%%%%%%%%%%%%%%%%%%%%%%%%%%%%%%%%%%%%%%%%%%%%%%%%%%%%%%%%%%%%%%%%%%%%%%%%%%%%%%%%%%%%%%%%%%
%\section{Experimental details and results}
%%%%%%%%%%%%%%%%%%%%%%%%%%%%%%%%%%%%%%%%%%%%%%%%%%%%%%%%%%%%%%%%%%%%%%%%%%%%%%%%%%%%%%%%%%%%%%%%%%%%%%%%%%%%%%%%%%%%%%%%%%%

To produce $^{214,216,218}$U nuclei, a series of experiments were performed at the gas-filled recoil separator, SHANS (Spectrometer for Heavy Atoms and Nuclear Structure) \cite{Zhang2013}, at the Heavy Ion Research Facility in Lanzhou (HIRFL), China. For $^{214}$U, the fusion-evaporation reaction of $^{182}$W($^{36}$Ar, 4n)$^{214}$U with a beam energy of 184~MeV and a typical beam intensity of $\sim$500~pnA was used. The $^{182}$W targets with a thickness of 300-350~$\mu$g/cm$^{2}$ were prepared by sputtering the material onto 80-$\mu$g/cm$^{2}$-thick carbon foils and then covered by 10-$\mu$g/cm$^{2}$-thick carbon layer. The recoiled evaporation residues (ERs) were separated efficiently by SHANS and collected by three 16-strip position-sensitive silicon detectors (PSSDs), which were mounted side by side at the focal plane of the separator. Each PSSD has an active area of 50$\times$50~mm$^{2}$. Due to the shallow implantation depth of $\sim$4~$\mu$m, the full-energy $\alpha$ particles emitted from ERs and/or their descendants were registered with an efficiency of $\sim$54\%. Eight side silicon detectors (SSDs) were mounted in front of the PSSDs in an open box geometry to measure the escaped $\alpha$ particles. For such events, the total $\alpha$-particle energy was reconstructed by adding the deposited energies in PSSD and SSD. In order to distinguish the $\alpha$-decay events from the implanted products, two multi-wire proportional counters were installed upstream from the PSSDs. A digital data readout system including waveform digitizers was used for the data acquisition. Details of the detection system and data analysis method were described in Refs.~\cite{Zhang2019, Ma2020, Yang2018, Yang2019}.

%%%%%%%%%%%%%%%%%%%%%%%%%%%%%%%%%%%%%%%%%%%%%%%%%%%%%%%%%%%%%%%%%%
\begin{figure}[!b]
  \includegraphics[width=0.5\textwidth]{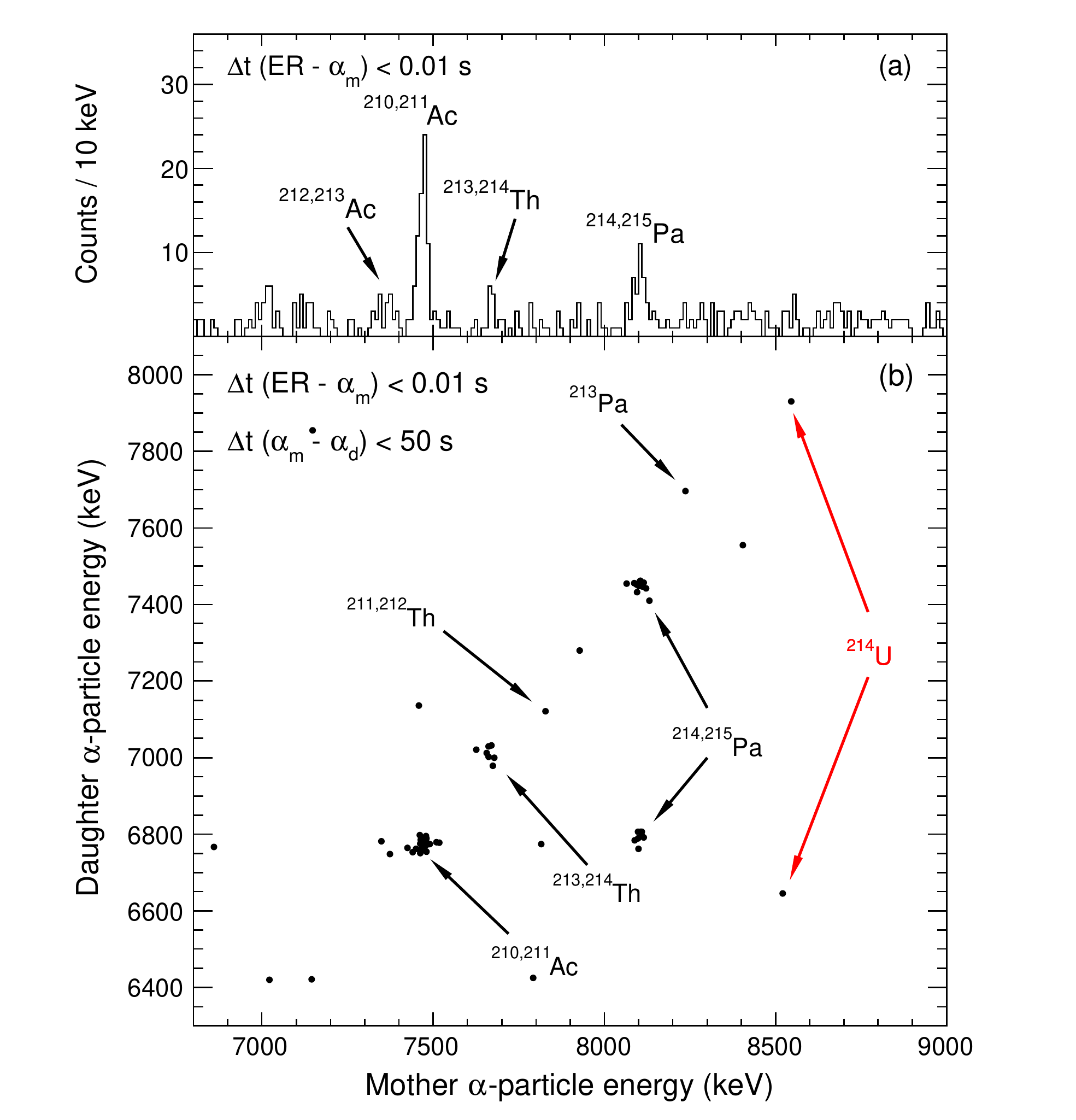}
  \caption{\label{fig1} a) Energy spectrum for $\alpha$-decay events following recoil implantations within a time window of 10 ms. b) Two-dimensional plot of mother and daughter $\alpha$-particle energies for $\text{ER}-\alpha_m-\alpha_d$ correlations in the $^{36}$Ar+$^{182}$W reaction. Maximum search times for the $\text{ER}-\alpha_m$ and $\alpha_m-\alpha_d$ pairs are 10 ms and 50 s, respectively. The decay events from the new isotope $^{214}$U are indicated by red arrows.}
\end{figure}
%%%%%%%%%%%%%%%%%%%%%%%%%%%%%%%%%%%%%%%%%%%%%%%%%%%%%%%%%%%%%%%%%%

\begin{figure}[!tb]
  \includegraphics[width=0.47\textwidth]{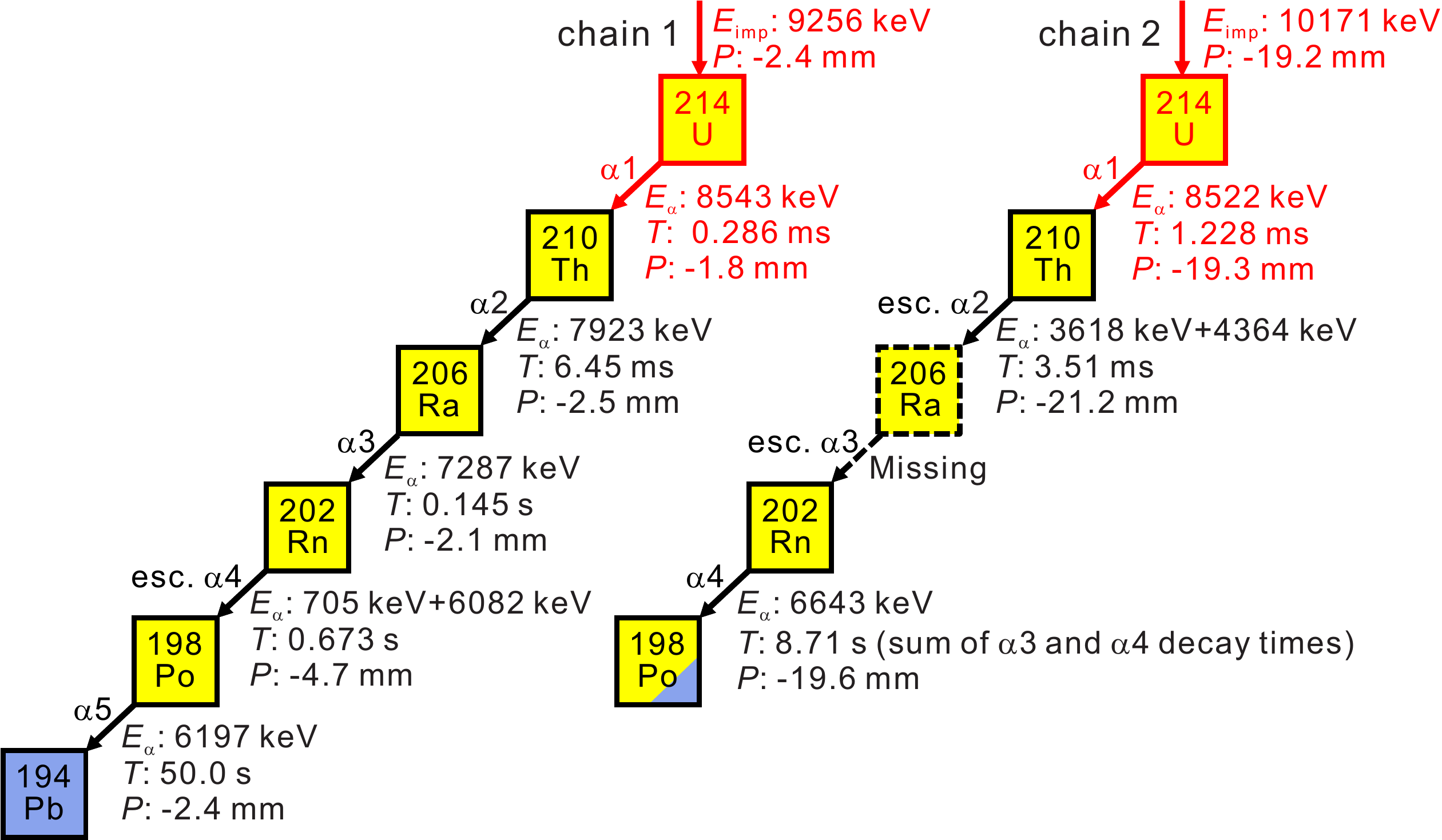}
  \caption{\label{fig2} Observed $\alpha$-decay chains for $^{214}$U. For each chain, the implantation energy of ERs ($E_{\rm{imp}}$), the $\alpha$-particle energy ($E_{\alpha}$), the decay time ($T$), and the position ($P$) in the strip detector are shown. The reconstructed energies for escaping $\alpha$ decays are given as the sum of the PSSD and SSD energies.}
\end{figure}

The identification of $^{214}$U was performed by searching for the position-time correlated $\alpha$-decay chains with the help of known $\alpha$-decay properties of its descendants. An energy spectrum for $\alpha$-decay events following the ERs and a two-dimensional plot for the decay energy correlation between mother and daughter nuclei ($\text{ER}-\alpha_m-\alpha_d$) are shown in Fig.~\ref{fig1}(a) and ~\ref{fig1}(b), respectively. The Pa, Th, and Ac isotopes were produced from charged-particle evaporation channels. Two decay events in Fig.~\ref{fig1}(b) were assigned to the new isotope $^{214}$U unambiguously. The details of these decay chains are displayed in Fig.~\ref{fig2}. The measured decay properties of daughter products match well with the known data \cite{NNDC} for $^{210}$Th, $^{206}$Ra, $^{202}$Rn, and $^{198}$Po. Based on these measurements, the mean $\alpha$-particle energy and half-life of $^{214}$U were determined to be 8533(18)~keV and $0.52^{+0.95}_{-0.21}$~ms, respectively, which are listed in Table \ref{tab1}. The uncertainties of half-life were estimated by the maximum likelihood method described in Ref.~\cite{Schmidt1984}. The production cross section for $^{214}$U was determined to be $10^{+14}_{-7}$ pb.

%%%%%%%%%%%%%%%%%%%%%%%%%%%%%%%%%%%%%%%%%%%%%%%%%%%%%%%%%%%%%%%%%%
%%%%%%%%%%%%%%%%%%%%%%%%%%%%%%%%%%%%%%%%%%%%%%%%%%%%%%%%%%%%%%%%%%

The properties of $^{216}$U, which was the lightest even-even uranium isotope known previously, were reported in our previous work \cite{Ma2015} and in Refs.~\cite{Devaraja2015, Wakabayashi2015}. However, at most four decay chains from the ground state of $^{216}$U were observed in each study, resulting in a relatively large uncertainty of decay half-life. In the present investigation, the same experimental setup as for $^{214}$U was used, but with a reaction of $^{180}$W($^{40}$Ar, 4n)$^{216}$U at a beam energy of 191~MeV. Thirteen decay chains were assigned to the ground-state-to-ground-state (g.s.-to-g.s.) decay of $^{216}$U. The deduced decay energy and half-life of $^{216\rm{g}}$U are 8374(17)~keV and $1.28^{+0.49}_{-0.28}$~ms, respectively. By combining all data from the present study and from Refs.~\cite{Ma2015, Devaraja2015, Wakabayashi2015}, the averaged half-life for the ground state of $^{216}$U was deduced to be $2.25^{+0.63}_{-0.40}$~ms. The results are compared with the literature data in Table \ref{tab1}.

In order to obtain more precise decay properties of $^{218}$U, two experiments with $^{182}$W($^{40}$Ar, 4n)$^{218}$U and $^{184}$W($^{40}$Ca, $\alpha$2n)$^{218}$U reactions were carried out with beam energies of 190~MeV and 206~MeV, respectively. Totally, 76 decay chains were assigned to the decay from the ground state of $^{218}$U, leading to the determination of $E_{\alpha}=8612(14)$~keV and $T_{1/2}=0.65^{+0.08}_{-0.07}$~ms. The uncertainties of half-life were improved in comparison with previous results \cite{Ma2015, Leppanen2007, Leppanen2005, Andreyev1992} (see Table \ref{tab1}).

%%%%%%%%%%%%%%%%%%%%%%%%%%%%%%%%%%%%%%%%%%%%%%%%%%%%%%%%%%%%%%%%%%
\squeezetable
\begin{table}[!b]
  \caption{\label{tab1} The g.s.-to-g.s.~$\alpha$-decay energies and half-lives of $^{214,216,218}$U measured in this work. The reduced $\alpha$-decay widths $\delta^2$, in column 4, are calculated by Rasmussen formalism \cite{Rasmussen1959} assuming the $\alpha$-particle angular momentum, $\Delta L = 0$. The data for $^{216,218}$U are compared with literature values.}
  \begin{ruledtabular}
  \begin{tabular}[t]{ccccccc}
    \multirow{2}{*}{Isotope} & \multicolumn{3}{c}{This work} & \multicolumn{3}{c}{Literature data} \\
    \cline{2-4} \cline{5-7}
    & $E_{\alpha}$/keV & $T_{1/2}$/ms & $\delta^{2}$/keV & $E_{\alpha}$/keV & $T_{1/2}$/ms & Ref. \\
    \hline
    $^{214}$U & 8533(18) & $0.52^{+0.95}_{-0.21}$ & $128^{+233}_{-52}$ & - & - & - \\

     & & & & & & \\
    \multirow{4}{*}{$^{216}$U}  & \multirow{4}{*}{8374(17)} & \multirow{4}{*}{$2.25^{+0.63}_{-0.40}$\footnotemark[1]} & \multirow{4}{*}{$78^{+22}_{-14}$}
     & 8384(30) & $4.72^{+4.72}_{-1.57}$ & \cite{Ma2015} \\
     & & & & 8340(50) & $3.8^{+8.8}_{-3.2}$ & \cite{Devaraja2015} \\
     & & & & 8390(33) & $2.6^{+3.6}_{-1.0}$ & \cite{Wakabayashi2015} \\

     & & & & & & \\
    \multirow{4}{*}{$^{218}$U}  & \multirow{4}{*}{8612(14)} & \multirow{4}{*}{$0.65^{+0.08}_{-0.07}$} & \multirow{4}{*}{$53^{+7}_{-6}$}
     & 8600(30) & $1.15^{+1.58}_{-0.42}$ & \cite{Ma2015} \\
     & & & & 8612(9) & $0.51^{+0.17}_{-0.10}$ & \cite{Leppanen2007, Leppanen2005} \\
     & & & & 8625(25) & $1.5^{+7.3}_{-0.7}$ & \cite{Andreyev1992} \\

  \end{tabular}
  \end{ruledtabular}

  \footnotetext[1]{The value is deduced by combining all 21 decay events from this work and Refs.~\cite{Ma2015, Devaraja2015, Wakabayashi2015}, and is also used for the decay width calculation for $^{216}$U.}
\end{table}
%%%%%%%%%%%%%%%%%%%%%%%%%%%%%%%%%%%%%%%%%%%%%%%%%%%%%%%%%%%%%%%%%%

%%%%%%%%%%%%%%%%%%%%%%%%%%%%%%%%%%%%%%%%%%%%%%%%%%%%%%%%%%%%%%%%%%%%%%%%%%%%%%%%%%%%%%%%%%%%%%%%%%%%%%%%%%%%%%%%%%%%%%%%%%%
%\section{Discussion}
%%%%%%%%%%%%%%%%%%%%%%%%%%%%%%%%%%%%%%%%%%%%%%%%%%%%%%%%%%%%%%%%%%%%%%%%%%%%%%%%%%%%%%%%%%%%%%%%%%%%%%%%%%%%%%%%%%%%%%%%%%%

\begin{figure}[!tb]
  \includegraphics[width=0.48\textwidth]{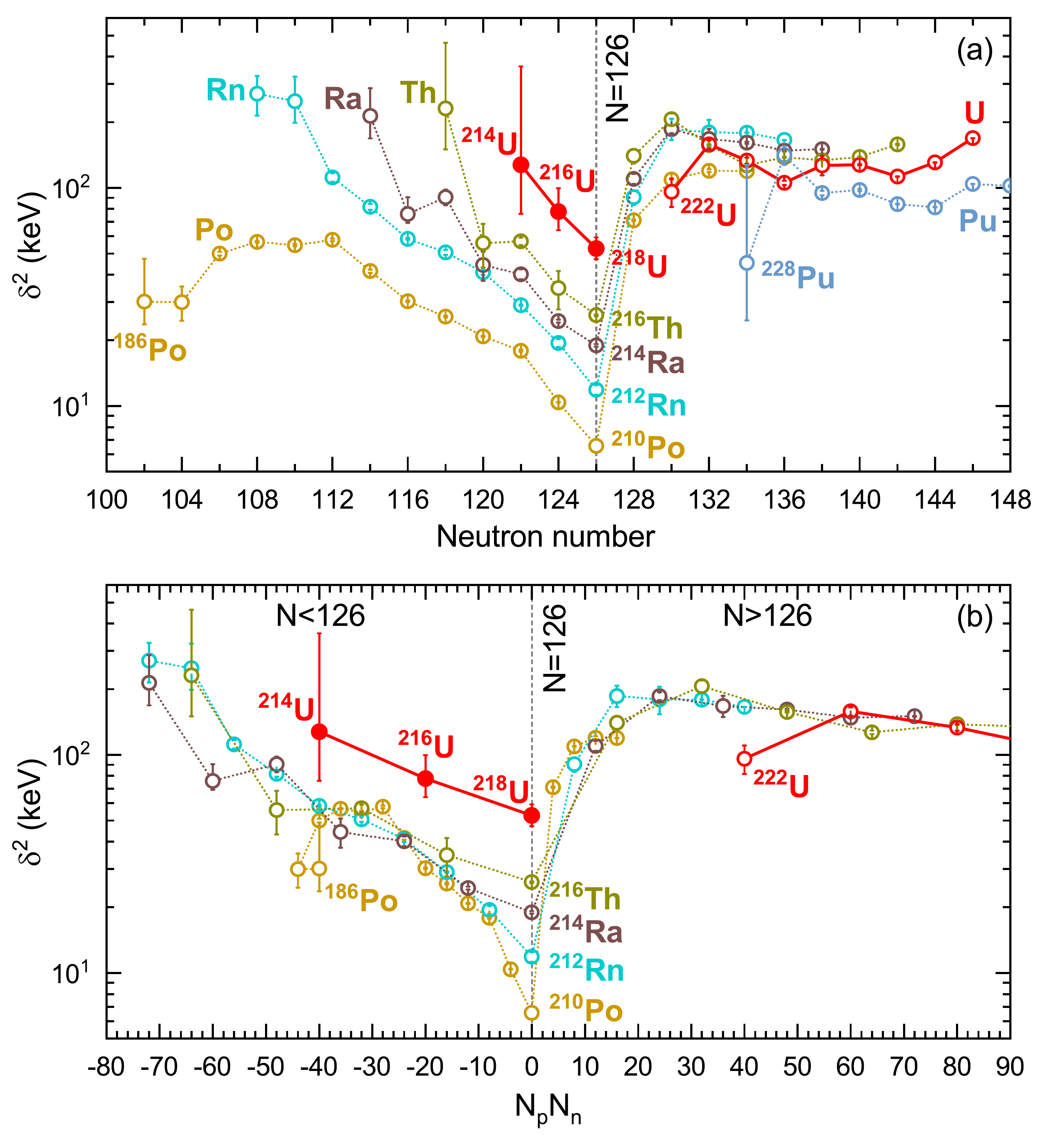}
  \caption{\label{fig3} (a) Systematics of reduced widths for g.s.-to-g.s.~$\alpha$ decays of even-even $84 \leq Z \leq 94$ isotopes as a function of neutron number. The decay properties are taken from Refs.~\cite{Andreyev2013, Khuyagbaatar2015, NNDC, Valli1968, Nishio2003, Heredia2010, Kuusiniemi2005}. The values for $^{214,216,218}$U from this work are shown by filled circles. The errors of reduced widths are only determined by half-life uncertainties. (b) Same as (a) but against $N_{p}N_{n}$ for even-even Po to U isotopes. The $N_p$ and $N_n$ values are calculated relative to $Z=82$ and $N=126$ closed shells, respectively, with an exception of $^{186}_{~84}$Po$_{102}$, for which $N_{n}=-20$, relative to the closest $N=82$ neutron shell.}
\end{figure}

To study the nuclear structure evolution in the $N=126$ region, the reduced widths $\delta^2$ for g.s.-to-g.s.~decays of even-even $84 \leq Z \leq 94$ nuclei are extracted by using Rasmussen method \cite{Rasmussen1959}, see Fig.~\ref{fig3}(a). The uncertainties of $\delta^2$ values are mostly due to the half-life uncertainties. The $^{214,216,218}$U values determined in this work are shown in column 4 of Table \ref{tab1} and plotted by filled circles in Fig.~\ref{fig3}(a).

%%%%%%%%%%%%%%%%%%%%%%%%%%%%%%%%%%%%%%%%%%%%%%%%%%%%%%%%%%%%%%%%%%
%%%%%%%%%%%%%%%%%%%%%%%%%%%%%%%%%%%%%%%%%%%%%%%%%%%%%%%%%%%%%%%%%%

In each of the Po, Rn, Ra, and Th isotopic chains, a sharp decrease of reduced widths at $N=126$ is well-established, indicating a notable neutron shell effect \cite{Rasmussen1959, Toth1984, Toth1986}. Our new data suggest for the first time that the minimum decay width for U isotopes is likely at $^{218}$U ($N=126$). This result is in contrast with our previous work \cite{Ma2015}, where only half a value of $\delta^2$($^{216}$U) ($34^{+34}_{-11}$~keV) was reported. A shrinking of the $\delta^2$ enhancement between the $N=126$ and $N=130$ isotones with the increasing of proton number was attributed to a weakening of the $N=126$ shell effect as suggested in Ref.~\cite{Khuyagbaatar2015}. The nearly constant or even decreasing values for the most neutron-deficient polonium isotopes were explained by the configuration mixing effect \cite{Andreyev2013, Duppen2018}.

In Fig.~\ref{fig3}(a), another important feature revealed by our new data is that, while the decay widths at $N=122$, 124, and 126 for Po--Th isotopes increase monotonously with increasing proton number, an unexpected sharp increase was observed from Th to U isotopes at the same neutron numbers. This suggests that the $\alpha$-particle formation probability is enhanced in these U isotopes.

In order to get a deeper insight into the behavior of reduced widths, we studied the influence of the $p$-$n$ interaction upon the $\alpha$-decay process in this mass region. Given the fact that the $N_{p}N_{n}$ scheme \cite{Casten1985a, Casten1985b} allows a uniform description of structure evolution for a variety of observables and highlights the importance of valence $p$-$n$ interaction \cite{Casten2001, Casten2016, Zhao2000, Seif2011, Bhattacharya2008, Sun2016}, the $\delta^2$ values are plotted against $N_{p}N_{n}$ in Fig.~\ref{fig3}(b). Here, $N_p$ and $N_n$ are the numbers of valence protons and neutrons relative to the nearest closed shells: $Z=82$ for proton and $N=126$ for neutron. It is striking to see that the $N_{p}N_{n}$ plot displays a remarkable simplification for the systematics of decay widths in this region. In the $N>126$ region, the $\delta^2$ values increase rapidly until $N_{p}N_{n}\approx20$, and then converge into a nearly constant value of $\sim$150~keV (except for $^{222}$U). This ``saturation'' phenomenon might indicate that the $\alpha$ decays in these nuclei are affected only slightly by the $p$-$n$ interaction, but are dominated by the $p$-$p$ and $n$-$n$ pairing interactions, as pointed out theoretically in Refs.~\cite{Tonozuka1979, Qi2010, Qi2019}. In other words, it is the strong pairing force among the protons and neutrons occupying high-$j$ orbits (e.g., $\pi 0h_{9/2}$ and $\nu 1g_{9/2}$) which leads to the large $\alpha$-particle formation probability \cite{Andreyev2013}.

In contrast, for the $N<126$ nuclei, the $\delta^2$ values for Po--Th isotopes show quite different behaviors, increasing exponentially with increasing the absolute $N_{p}N_{n}$ quantity along a relatively compact tendency (except for $^{186,188}$Po). The increasing trend can be partly explained by the increasing neutron and proton pairing correlations \cite{Andreyev2013} as one moves away from $N=126$ and $Z=82$. More importantly, considering that the $N_{p}N_{n}$ value provides a reliable measure of interaction between the valance protons and neutrons \cite{Casten2001, Casten2016}, this specific feature shown in Fig.~\ref{fig3}(b) implies that the $p$-$n$ interaction can also play an essential role in the $\alpha$-particle clustering in this region.

%%%%%%%%%%%%%%%%%%%%%%%%%%%%%%%%%%%%%%%%%%%%%%%%%%%%%%%%%%%%%%%%%%
\begin{figure}[!tb]
  \includegraphics[width=0.48\textwidth]{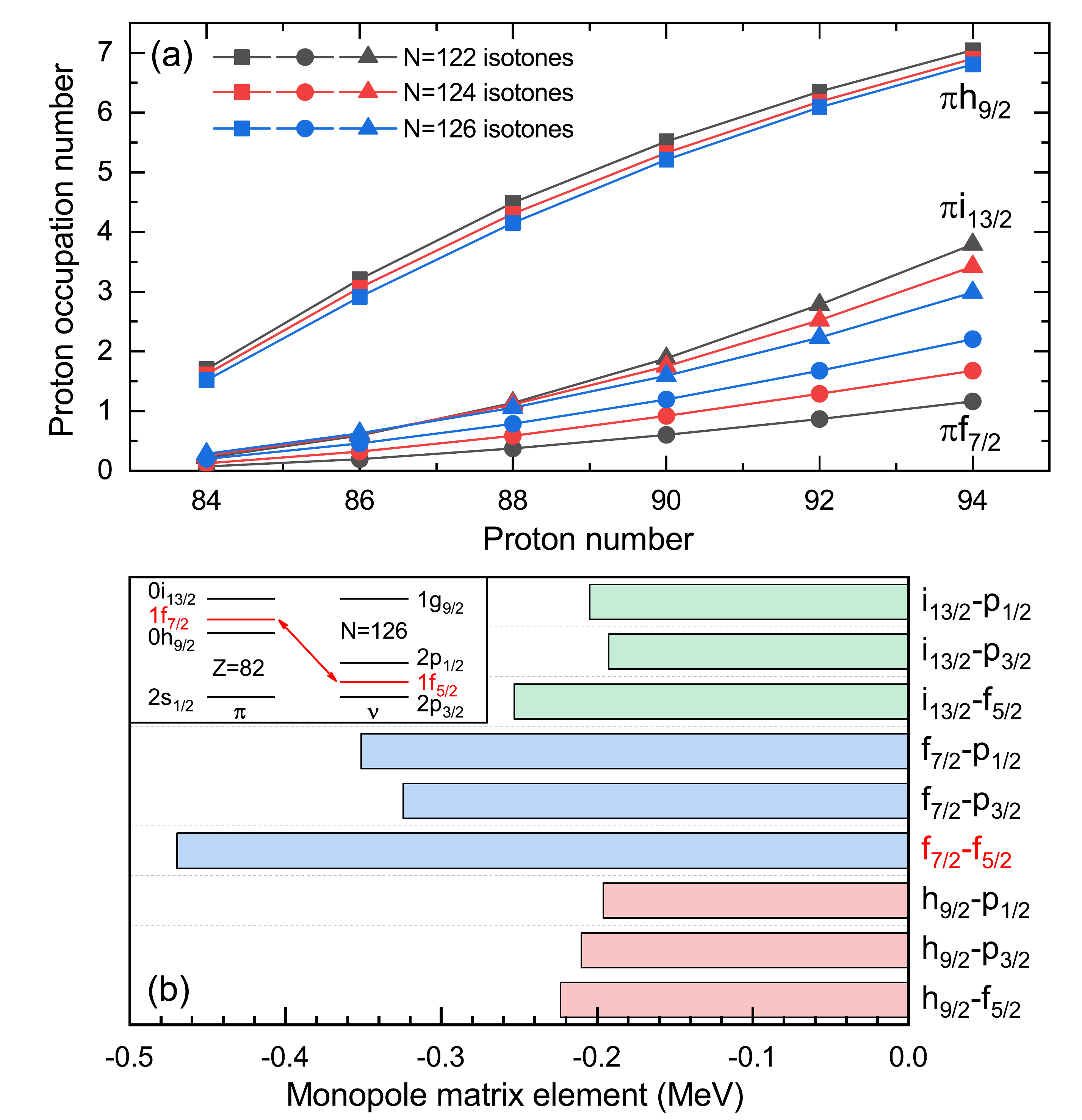}
  \caption{\label{fig4} (a) Calculated proton occupation numbers for the $\pi 0h_{9/2}$ (square), $\pi 1f_{7/2}$ (circle), and $\pi 0i_{13/2}$ (triangle) orbits in $N=122$, 124, and 126 even-$Z$ isotones of Po--Pu. (b) Monopole matrix elements of $p$-$n$ interaction calculated for the $Z>82$ and $N\leq126$ nuclei. The inset in panel (b) shows the single-particle orbits near the $^{208}$Pb doubly closed shells, and the strong interaction between the $1f_{7/2}$ protons and $1f_{5/2}$ neutrons is marked.}
\end{figure}
%%%%%%%%%%%%%%%%%%%%%%%%%%%%%%%%%%%%%%%%%%%%%%%%%%%%%%%%%%%%%%%%%%

The $\delta^2$ values of $^{214,216}$U, however, show striking discrepancy with the unified trend established for $84 \leq Z \leq 90$ and $N<126$ nuclei. Regardless the relatively large uncertainties for $^{214}$U, a significant enhancement by a factor of two is revealed for $^{214,216}$U as shown in Fig.~\ref{fig3}(b). This new feature might be related to the possible changes of occupancy of $0h_{9/2}$ and $1f_{7/2}$ proton orbits approaching $Z=92$. Indeed, below $Z=92$, it is expected that the $0h_{9/2}$ protons play a dominant role, which is confirmed by, e.g., the $9/2^-$ ground states for most of odd-$A$ $_{83}$Bi, $_{85}$At, $_{87}$Fr, and $_{89}$Ac isotopes \cite{NNDC}. The $0h_{9/2}$ orbit is expected to be highly occupied in U ($Z=92$) with an enhanced probability of proton occupancy of the higher-lying $1f_{7/2}$ orbit. The later, combined with the neutron occupancy of $1f_{5/2}$ orbit around $N=118$--124, might lead to a strong monopole $p$-$n$ interaction (see inset of Fig.~\ref{fig4}(b)), which enhances the preformation probability in $\alpha$ decay.

In order to verify this conjecture, we have performed large-scale shell model calculations for the $84 \leq Z \leq 94$ and $N=122$, 124, 126 even-even nuclei. The same model spaces with the $0h_{9/2}$, $1f_{7/2}$, $0i_{13/2}$, $2p_{3/2}$, $1f_{5/2}$, and $2p_{1/2}$ orbits were selected for protons and neutrons. The single-particle energies are  fixed to those of $^{209}$Bi and $^{207}$Pb. The $p$-$p$, $n$-$n$, and $p$-$n$ parts of two-body interactions are taken from the Kuo-Herling particle interaction \cite{Warburton1991a}, Kuo-Herling hole interaction \cite{Warburton1991b}, and monopole based universal interaction \cite{Otsuka2010} plus M3Y spin-orbit interaction \cite{Bertsch1977}, respectively. Given the computational limit, the restrictions, for which the $\pi$($2p_{3/2}$, $1f_{5/2}$, $2p_{1/2}$) orbits are fully empty for protons and the $\nu$($0h_{9/2}$, $1f_{7/2}$, $0i_{13/2}$) orbits are fully occupied for neutrons, were made. The calculated proton occupation numbers for the $\pi 0h_{9/2}$, $\pi 1f_{7/2}$, and $\pi 0i_{13/2}$ orbits in $N=122$, 124, and 126 even-$Z$ isotones are shown in Fig.~\ref{fig4}(a). It can be seen that, due to the pairing correlation effect \cite{Casten2001}, the valence protons occupy mainly the $0h_{9/2}$ orbit with the increasing occupation probability of the $1f_{7/2}$ and $0i_{13/2}$ protons from Po to Pu isotopes. In particular, the effective proton occupation numbers for the $1f_{7/2}$ orbit in U and Pu isotopes are almost equal to or even higher than one.

The calculated monopole matrix elements between the proton and neutron orbits for the $Z>82$ and $N \leq 126$ nuclei are shown in Fig.~\ref{fig4}(b). The calculations demonstrate that all the $p$-$n$ interactions involving $1f_{7/2}$ protons are about twice more attractive than those involving $0h_{9/2}$ and $0i_{13/2}$ protons. In particular, the $\pi 1f_{7/2}$--$\nu 1f_{5/2}$ interaction is by far the strongest one in this region of nuclei. Therefore, the strong $p$-$n$ interactions related to the $1f_{7/2}$ protons, together with the increased occupancy of the $\pi 1f_{7/2}$ orbit, would lead to the enhanced $\alpha$-particle formation probability in the $N=122$, 124, and 126 uranium isotopes.

%%%%%%%%%%%%%%%%%%%%%%%%%%%%%%%%%%%%%%%%%%%%%%%%%%%%%%%%%%%%%%%%%%%%%%%%%%%%%%%%%%%%%%%%%%%%%%%%%%%%%%%%%%%%%%%%%%%%%%%%%%%
%\section{Conclusion}
%%%%%%%%%%%%%%%%%%%%%%%%%%%%%%%%%%%%%%%%%%%%%%%%%%%%%%%%%%%%%%%%%%%%%%%%%%%%%%%%%%%%%%%%%%%%%%%%%%%%%%%%%%%%%%%%%%%%%%%%%%%

In summary, a new isotope $^{214}$U was identified and improved $\alpha$-decay properties of $^{216,218}$U were measured by employing the gas-filled recoil separator SHANS and recoil-$\alpha$ correlation method. By combining the new and previously known data, we extracted the $\alpha$-decay reduced widths $\delta^2$ for the even-even Po--Pu nuclei with Rasmussen method. It is found that the $\delta^2$ systematics from Po to Th can be merged into two compact trends for the $N<126$ and $N>126$ nuclei in terms of $N_{p}N_{n}$ scheme. The behavior in the $N<126$ region indicates a crucial role played by $p$-$n$ interaction in $\alpha$ decay. Meanwhile, it is strikingly found that the reduced widths of $^{214,216}$U are enhanced remarkably by a factor of two relative to the systematic trend of $N<126$ nuclei in the $N_{p}N_{n}$ scheme. This might be explained as being due to the strong monopole interaction between the valence $1f_{7/2}$ protons and $1f_{5/2}$ neutrons combined with increased occupancy of the $f_{7/2}$ proton orbit, which was confirmed by the large-scale shell model calculations.

As a possible outlook for the future studies in this region, it is expected that, in view of the continuously increasing proton occupancy of the $1f_{7/2}$ orbit and the further enhancement of $p$-$n$ interaction, this effect might become even stronger in the Pu isotopes. Thus, it is extremely intriguing to extend the $\delta^2$ systematics to higher-$Z$ nuclei.
\\

\begin{acknowledgments}
The authors would like to thank the accelerator crew of HIRFL for providing the stable beams. This work was supported by the Strategic Priority Research Program of Chinese Academy of Sciences (Grant No.~XDB34010000), the National Key R\&D Program of China (Contract No.~2018YFA0404402), the National Natural Science Foundation of China (Grants No.~U1732270, No.~11975279, No.~11775316, No.~U1932139, No.~11961141004, No.~11735017, No.~12035011, No.~11965003, No.~11675225, No.~11635003, No.~11961141004, No.~U1867212, No.~11805289, No.~U1732139), the Chinese Academy of Sciences (Grant No.~QYZDJ-SSW-SLH041), the Youth Innovation Promotion Association CAS (2020409, 2017456), the Natural Science Foundation of Guangxi (Grants No.~2017GXNSFAA198160 and No.~2017GXNSFGA198001), and the Science Technology Facility Council (UK).
\end{acknowledgments}

%%%%%%%%%%%%%%%%%%%%%%%%%%%%%%%%%%%%%%%%%%%%%%%%%%%%%%%%%%%%%%%%%%%%%%%%%%%%%%%%%%%%%%%%%%%%%%%%%%%%%%%%%%%%%%%%%%%%%%%%%%%
%%%%%%%%%%%%%%%%%%%%%%%%%%%%%%%%%%%%%%%%%%%%%%%%%%%%%%%%%%%%%%%%%%%%%%%%%%%%%%%%%%%%%%%%%%%%%%%%%%%%%%%%%%%%%%%%%%%%%%%%%%%

\bibliography{214U}

\end{document}